\documentclass[fleqn]{llncs}
\usepackage[utf8]{inputenc}
\usepackage[T1]{fontenc}

\usepackage[scaled=0.8]{beramono}

\usepackage{amsmath}
\usepackage{amssymb}
\usepackage{mathtools}
\usepackage{proof}

\usepackage{xspace}

\usepackage{syntax}
\usepackage[numbers]{natbib}

\usepackage{hyperref}
\hypersetup{hidelinks,
    colorlinks=true,
    allcolors=blue,
    pdfstartview=Fit,
    breaklinks=true}

\usepackage[capitalise,nameinlink]{cleveref}

\newcommand{\Cuvee}{\textsc{Cuvée}\xspace}
\newcommand{\code}[1]{\texttt{#1}}
\newcommand{\hoare}[3]{\{ #1 \}~#2~\{ #3 \}}
\newcommand{\hoaretotal}[3]{[ #1 ]~#2~[ #3 ]}

\newcommand{\xs}{\underline x}
\newcommand{\ys}{\underline y}

\newcommand{\nonterminal}[1]{\ensuremath{\langle \mathit{#1} \rangle}}

\title{\Cuvee: Blending SMT-LIB with \\ Programs and Weakest Preconditions}
\author{Gidon Ernst}
\institute{LMU Munich, Germany, \email{gidon.ernst@lmu.de}}

\begin{document}
\maketitle

\begin{abstract}
% \url{https://github.com/gernst/cuvee/blob/master/ae.zip}\\
\Cuvee is a program verification tool that reads SMT-LIB-like input files where terms may additionally contain weakest precondition operators over abstract programs.
\Cuvee translates such inputs into first-order SMT-LIB by symbolically executing these programs.
The input format used by \Cuvee is intended to achieve a similar unification of tools for that for example synthesize loop summaries.
A notable technical aspect of \Cuvee itself is the consequent use of loop pre-/postconditions instead of invariants, and we demonstrate how this lowers the annotation burden on some simple while programs.
% Moreover, \Cuvee can express refinement conditions between multiple programs.
\end{abstract}


\section{Introduction}

Intermediate verification languages and tools such as Boogie~\cite{leino2008boogie}, Why3~\cite{bobot2011why3}, and Viper~\cite{muller2016viper}
have had a significant impact on the state-of-the-art of (deductive) program verification.
At the annual competition on interactive program verification VerifyThis~\cite{ernst2019verifythis},
tools like these are put to practice on small but intricate verification problems.
Currently, VerifyThis lacks a common input format.
This is fine for hard challenges where efficiency of the tool and seamless interaction are required.
SMT-LIB~\cite{barrett2010smt} is a standardized interchange format for verification tasks in first-order logic
that is widely used in many different application domains such as constraint-solving, program verification, and model-checking.
Many mature tools are available, and the annual SMT-COMP evaluates and compares their performance on benchmark problems.
Part of the success of the SMT-LIB format is its regular syntax and its precise standardization.
SV-COMP~\cite{beyer2020advances} is a competition where the participating tools analyze C code fully automatically.
It is therefore restricted to less complex properties that cannot be solved without the guidance of a proof engineer.

The goal of this work is to occupy the spot between these communities,
as well as the different technologies involved~\cite{bartocci2019toolympics}.
How to make use, e.g., of tools that can infer invariants?
The input format that is described here and implemented in \Cuvee takes deliberate trade-offs
to position itself between the highly expressive logic of e.g. Dafny,
and stressing the interoperability that made SMT-LIB successful.
As such one design goal, for example, is to re-use the existing logical data types of that format,
even if it might be less convenient as a front-end for humans.

\section{Syntax}
\label{sec:syntax}

The new constructs accepted by \Cuvee are shown in \cref{fig:syntax}.

\begin{figure}[t]
    \begin{grammar}
    <var_binding>
        ::= "(" <symbol> <term> ")"

    <term>
        ::=  "(wp" <program> <term>")"
        \alt "(box" <program> <term>")"
        \alt "(dia" <program> <term>")"
        \alt "(old" <term> ")"
        \alt ...

    <program>
        ::=  "(assign" "("<var_binding>")"$^+$ ")"
        \alt "(spec"   "(" <symbol>$^+$ ")" <term> <term> ")"
        \alt "(block"  <program>$^+$ ")"
        \alt "(if"     <term> <term> <term> ")"
        \alt "(while"  <term> <term> <attribute>$^+$ ")"

    <command>
        ::=  "(assert-counterexample" <term> <program> <term> ")"
        \alt ...
    \end{grammar}
    \caption{Extension of the SMT-LIB supported by \Cuvee:
             Weakest-precondition operators,
             simple nondeterministic \textsc{While} programs,
             and a top-level command to specify Hoare triples
             (\nonterminal{var\_binding} is from the SMT-LIB grammar).}
    \label{fig:syntax}
\end{figure}

Terms \nonterminal{term} of SMT-LIB are extended by three weakest-precondition operators,
\code{(wp  $p$ $t$)}, \code{(box $p$ $t$)}, \code{(dia $p$ $t$)}
for programs~$p$ (see below)
and postconditions~$t$ (terms of sort \code{Bool},
which state slightly different correctness criteria with respect to termination and nondeterministic choices.

The first one, \code{wp} denotes Dijkstra's well-known weakest precondition:
All executions of~$p$, when started in the current state, terminate and lead to a state that satisfies~$t$.
For instance, a Hoare triple for total correctness $\hoaretotal{\phi}{p}{\psi}$ can be written as the implication \code{(=> $\phi$ (wp $p$ $\psi$))}

Similarly, \code{box} does not require termination (i.e., it expresses the weakest liberal precondition),
i.e., $\hoare{\phi}{p}{\psi}$ can be written as \code{(=> $\phi$ (box $p$ $\psi$))}.

The operator \code{dia} reflects angelic execution instead of demonic execution:
For \code{(dia $p$ $t$)} it is required that there is at least one execution of~$p$
that terminates and leads to a state that satisfies~$t$.
The names of the latter two operators is taken from Dynamic Logic.

The expression \code{(old $t$)} is intended for use in loop annotations.
It refers to previous states of the execution at the beginning of loop iterations.

\medskip

Programs \nonterminal{program} provide familiar constructs from a simple sequential \textsc{While} language.

Parallel assignments evaluate all right-hand-sides simultaneously,
e.g. \code{(assign (x y)  (y x))} swaps the values stored in the variables \code{x} and \code{y}.
Note that there is no syntactic difference between program variables and logical ones,
and the former ones may range over arbitrary SMT-LIB data types.

Specification statements \code{(spec ($x_1 \cdots x_n$) $\phi \psi$)} (see~\cite{morgan1988specification})
encode arbitrary, possibly nondeeterministic transitions.
The effect of executing such a statement is that the precondition~$\phi$ of the statement
is asserted (i.e., needs to hold in the current state),
then the variables \code{$x_1 \cdots x_n$} are given fresh arbitrary values,
and the postcondition $\psi$ is assumed for the remainder of the execution.

Specification statements can encode assertions (\code{assert $\phi$} becomes \code{(spec () $\phi$ true)}),
assumptions (\code{assume $\psi$} becomes \code{(spec () true $\psi$)}),
and the statement \code{havoc $x_1 \cdots x_n$} from e.g. Boogie
(which becomes \code{(spec ($x_1 \cdots x_n$) true true)}).

Within the postcondition of a specification statement, \code{old} refers to the pre-state of the statement itself.
This admits elegant encoding of transition relations,
i.e., \code{(spec (x) true (> x (old x)))} specifies that the new value of \code{x} is strictly larger than the previous one.

Specification statements are useful internally, too, to encode the inductive hypothesis of the loop rule implemented in \Cuvee (see \cref{sec:syntax}).

Sequential composition is written as \code{(block $p_1 \cdots p_n$)}, where \code{(block)} denotes the empty statement.
Conditional statements \code{(if $b$ $p_1$ $p_2$)} execute either~$p_1$ or~$p_2$ depending of the evaluation of the test~$b$ (a boolean term) in the current state.

\medskip
\noindent
While loops
\[ \code{(while $b$ $p$ :termination $t$ :precondition $\phi$ :postcondition $\psi$)} \]
execute $p$ as long as the test~$b$ holds true.
\Cuvee supports some attributes that can be used to specify loop annotations,
namely, a termination measure~$t$, a loop precondition~$\phi$, and a loop postcondition~$\psi$.
All three annotations are optional.

\medskip

A new top-level command \code{(assert-counterexample $\phi$ $p$ $\psi$)}
asserts that the Hoare triple $\hoaretotal{\phi}{p}{\psi}$ is not valid.
This command roughly translates to \code{(assert (not (=> $\phi$ (wp $p$ $\psi$))))},
however, \Cuvee implements special cases when $p$ is a while loop without a pre-/postcondition annotation
to derive the loop specification from such a contract.
Moreover, within a \code{assert-counterexample} command,
\code{old} in the postcondition $\psi$ refers to the pre-state.
Note that this feature is currently not expressible in the expression/program language alone,
but can be emulated if needed by introducing additional logical variables capturing this pre-state explicitly.

Consistent with the standard pattern in SMT-LIB, where formulas are proved by searching for satisfying assignments of the negation,
an \code{unsat} from the underlying SMT solver on the problem translated by \Cuvee indicates that there is no counterexample
i.e., the program is correct with respect to the specified contract.

\section{Proof Rules}
\label{sec:semantics}

\begin{figure}[t]
    \begin{align*}
\code{(wp (assign (($x_1$ $t_1$) $\cdots$ ($x_n$ $t_n$))) $Q$)}
    & \equiv Q[x_1,\ldots,x_n \gets t_1,\ldots,t_n]
    \\
\code{(wp (block $p_1 \cdots p_n$ $Q$)}
    & \equiv \code{(wp $p_1$ $\cdots$ (wp $p_n$ $Q$))}
    \\
\code{(wp (if $b$ $p_1$ $p_2$) $Q$)}
    & \equiv        \code{(and (=> $b$ (wp $p_1$ $Q$))} \\
    & \hspace{30pt} \code{     (=> (not $b$) (wp $p_2$ $Q$)))}
    \end{align*}

    \begin{gather*}
\infer{\code{(wp (spec ($\xs$) $\phi$ $\psi$) $Q$)}}
{   \phi
 && \code{(forall ($\ys$) $\psi[\underline{\code{(old $x$)}} \gets \xs, \xs \gets \ys]
                            \to Q[\xs \gets \ys]$)}
              }
    \\[1em]
\infer{\code{(box (while $b$ $p$) $Q$)}}
{   
\begin{array}{c}
   \code{(box (spec ($\xs$) $\phi$ $\psi$) $Q$)} \\
   \code{(forall ($\xs$) (=> (and $\phi$ (not $b$)) $Q[\underline{\code{(old $x$)}} \gets \xs]$))} \\
   \code{(forall ($\xs$) (=> (and $\phi$ $b$) (box (block $p$ (spec ($\xs$) $\phi$ $\psi$)) $\psi$)$[\underline{\code{(old $x$)}} \gets \xs]$))}
\end{array}}
    \\
        \qquad \text{where $\phi$ is the precondition of the loop and $\psi$ is the postcondition,} \\
        \qquad \text{$\xs$ are the variables modified by the loop body~$p$ and $\ys$ are fresh.}
    \end{gather*}
    \caption{Predicate transformer semantics for the programming language in \cref{fig:syntax}.
             The rules for the specification statement and loops are given as inference rules
             to improve readability (these could be equally formulated as equivalences).}
    \label{fig:semantics}
\end{figure}

The predicate transformer semantics~\cite{manes2004predicate} of the input language is shown in \cref{fig:semantics} for the \code{wp} operator,
with the exception of the loop rule, which is shortened to omit the termination conditions
and therefore expressed using \code{box}. The rules for \code{box} and \code{dia} are similar.
Note that~$Q$ is a term that may again contain weakest-precondition operators.

The first three equivalences are standard.
Assignments propagate as a simultaneous substitution into the postcondition~$Q$.
Sequential execution simply nests the weakest precondition operator.
Conditionals produce two branches that evaluate the test positively resp.~negatively.

The specification statement binds fresh copies~$\ys$ for the havoc'ed variables~$\xs$,
and substitutes \code{old} expressions in the postcondition~$\psi$.
The first premise simply asserts the precondition~$\phi$.

\medskip

The loop rule is more involved and proceeds by induction on the number of loop iterations,
see Alexandru's bachelor thesis~\cite{alexandru2019}.
It has three premises:
The first, abstracts the entire execution of the loop loop with a specification statement,
after which the postcondition~$Q$ needs to hold.

The second premise corresponds to the base case of the induction
and establishes $Q$ upon termination (when $b$ is false).
Old values in~$Q$ refer to the current state.

The third premise corresponds to the inductive case, where~$b$ holds.
Then, the condition to show is that after executing the loop body~$p$ once,
the inductive hypothesis about the remaining iterations is sufficient to establish~$\psi$ after the entire loop.
This hypothesis is encoded as a specification statement
\code{(spec ($\xs$) $\phi$ $\psi$)}
that abstracts the remaining iterations, similarly as in the first premise.

The interaction between the different occurrences of \code{old} is somewhat intricate:
Within the specification statement, \code{old} refers to the state \emph{after} the first iteration,
according to the corresponding proof rule.
Hence, the it encodes that it is possible at that point to turn the precondition~$\phi$
into an arbitrary state after the remaining iterations that satisfies~$\psi$ for fresh copies of the modified variables~$\xs$
introduced by the specification statement.
Ultimately, this knowledge needs to suffice to establish~$\psi$ after the loop with \emph{all} iterations,
where \code{old} refers to the state \emph{before} executing the first iteration~$p$.

\section{Tool Description}
\label{sec:tool}

\Cuvee is implemented in the Scala programming language%
    \footnote{\url{https://scala-lang.org}}
and relies on SMT solvers as back-ends to solve the first-order verification conditions.

\Cuvee is open source under the MIT License at \code{https://github.com/gernst/cuvee}.

It reads one or more input files, reduces the weakest-precondition operators according to the rules in~\cref{fig:semantics}
using symbolic execution (i.e., the substitutions are delayed and propagated down the term structure).

The tool can be invoked from the command line in different ways
as exemplified below:
\begin{verbatim}
./cuvee                                 # read from stdin, write to stdin
./cuvee <file> -o <out>                 # read from file,  write to out
./cuvee <file1> ... <filen> -- ./z3 -in # invoke SMT solver directly
\end{verbatim}
It can either save the generated SMT-LIB script to a file,
or invoke an SMT solver, whose command line is appended after \verb|--|, Z3 in this case.
In this mode of operation, it pipes the generated verification task directly to the solver.
There are builtin abbreviations \verb|-z3| and \verb|-cvc4| that pass the necessary arguments,
assuming those solvers are present in \verb|$PATH|.

\section{Example}

\begin{figure}[t]
\begin{verbatim}
(declare-const x Int)
(declare-const y Int)
(declare-const a (Array Int Int))

(assert-counterexample
  (<= x y)
  (while (not (= x y))
         (if (<= (select a x) (select a y))
             (assign (x (+ x 1)))
             (assign (y (- y 1))))
         :termination (- y x))
  (forall ((z Int))
    (=> (and (<= (old x) z)
             (<= z (old y)))
        (<= (select a z) (select a x)))))

(check-sat)
\end{verbatim}
\caption{Finding the maximum in an array by elimination.}
\label{fig:example}
\end{figure}

We demonstrate \Cuvee on a simple example, shown in \cref{fig:example},
which is taken from the VerifyThis competition~2012.
It is an algorithm that searches the maximum element in an array by elimination.
It maintains two indices, $\code{x} \le \code{y}$
and moves the one pointing to the smaller element in each iteration.
The postcondition asserts that for any valid index \code{z},
the returned index \code{x} contains an element that is in fact at least as large.

\Cuvee infers the precondition and postcondition of the loop from the
specified contract, such that the example is solved without further interaction.

A few more examples are contained in the \code{examples} subfolder on github,
including GCD, and mapping an array range by an unspecified function.

\section{Conclusion and Outlook}

\Cuvee is a tool that aims to bridge the gap between fully automated program verification
and approaches that have the human in the loop.
It does so by extending SMT-LIB to cover weakest precondition statements about abstract programs,
which tightly integrate into the existing standard.

There are plenty of opportunities for future development.
For one, it is hoped that the format will be taken up by others,
supported by standardization efforts.
Extension to recursive procedures is planned as well.

Ultimately, following this approach opens up the possibility to evaluate and compare
verification tools that can handle complex functional correctness conditions
involving quantifiers, arrays, and other data types.

\bibliographystyle{splncsnat}
\bibliography{references}

\begin{thebibliography}{10}
\providecommand{\natexlab}[1]{#1}
\providecommand{\url}[1]{\texttt{#1}}
\providecommand{\urlprefix}{}

\bibitem[{Alexandru(2019)}]{alexandru2019}
Alexandru, G.: Specifying loops with contracts (2019), {Bachelor's Thesis, LMU
  Munich}

\bibitem[{Barrett et~al.(2010)Barrett, Stump, Tinelli et~al.}]{barrett2010smt}
Barrett, C., Stump, A., Tinelli, C., et~al.: The smt-lib standard: Version 2.0.
\newblock In: Proceedings of the 8th International Workshop on Satisfiability
  Modulo Theories (Edinburgh, England). vol.~13, p.~14 (2010)

\bibitem[{Bartocci et~al.(2019)Bartocci, Beyer, Black, Fedyukovich, Garavel,
  Hartmanns, Huisman, Kordon, Nagele, Sighireanu
  et~al.}]{bartocci2019toolympics}
Bartocci, E., Beyer, D., Black, P.E., Fedyukovich, G., Garavel, H., Hartmanns,
  A., Huisman, M., Kordon, F., Nagele, J., Sighireanu, M., et~al.: Toolympics
  2019: an overview of competitions in formal methods.
\newblock In: International Conference on Tools and Algorithms for the
  Construction and Analysis of Systems. pp. 3--24. Springer (2019)

\bibitem[{Beyer(2020)}]{beyer2020advances}
Beyer, D.: Advances in automatic software verification: Sv-comp 2020.
\newblock In: International Conference on Tools and Algorithms for the
  Construction and Analysis of Systems. pp. 347--367. Springer (2020)

\bibitem[{Bobot et~al.(2011)Bobot, Filli{\^a}tre, March{\'e}, and
  Paskevich}]{bobot2011why3}
Bobot, F., Filli{\^a}tre, J.C., March{\'e}, C., Paskevich, A.: Why3: Shepherd
  your herd of provers.
\newblock In: Boogie 2011: First International Workshop on Intermediate
  Verification Languages. pp. 53--64 (2011)

\bibitem[{Ernst et~al.(2019)Ernst, Huisman, Mostowski, and
  Ulbrich}]{ernst2019verifythis}
Ernst, G., Huisman, M., Mostowski, W., Ulbrich, M.: {VerifyThis}--verification
  competition with a human factor.
\newblock In: International Conference on Tools and Algorithms for the
  Construction and Analysis of Systems. pp. 176--195. Springer (2019)

\bibitem[{Leino(2008)}]{leino2008boogie}
Leino, K.R.M.: This is {Boogie}~2 (2008), microsoft RiSE

\bibitem[{Manes(2004)}]{manes2004predicate}
Manes, E.G.: Predicate transformer semantics, vol.~33.
\newblock Cambridge University Press (2004)

\bibitem[{Morgan(1988)}]{morgan1988specification}
Morgan, C.: The specification statement.
\newblock ACM Transactions on Programming Languages and Systems (TOPLAS) 10(3),
  403--419 (1988)

\bibitem[{M{\"u}ller et~al.(2016)M{\"u}ller, Schwerhoff, and
  Summers}]{muller2016viper}
M{\"u}ller, P., Schwerhoff, M., Summers, A.J.: Viper: A verification
  infrastructure for permission-based reasoning.
\newblock In: International Conference on Verification, Model Checking, and
  Abstract Interpretation. pp. 41--62. Springer (2016)

\end{thebibliography}

\end{document}